\documentstyle[aps]{revtex}
\begin{document}
\title{Quantum superluminal communication does not result in the causal loop}
\author{Gao shan}
\address{Institute of Quantum Mechanics, 11-10, NO.10 Building,
YueTan XiJie DongLi, XiCheng District, Beijing 100045, P.R.China.
Tel: 86 010 82882274  Fax: 86 010 82882019 Email:
gaoshan.iqm@263.net }
\date{June 14, 1999}
\maketitle

\vskip 0.5cm

\begin{abstract}
We show that the quantum superluminal communication based on the
quantum nonlocal influence, if exists, will not result in the
causal loop, this conclusion is essentially determined by the
peculiarity of the quantum nonlocal influence itself, according to
which there must exist a preferred Lorentz frame for consistently
describing the quantum nonlocal process.
\end{abstract}

\vskip 0.2cm

As we say quantum mechanics permits no superluminal
communication\cite{QSC1}, we should refer to the present quantum
theory, and realize that the concrete reason is not related to the
peculiarity of the quantum nonlocal influence, which is manifested
in Bell's theorem\cite{Bell} and has been confirmed by more and
more experiments\cite{Aspect}, on the contrary, this kind of
quantum nonlocal influence may help to achieve the superluminal
communication when transcending the present quantum theory.

But, on the other hand, people may naturally argue that even
regardless of the limitation from present quantum theory, special
relativity will also inhibit such superluminal communication based
on the quantum nonlocal influence owing to the causal loop, thus
superluminal communication is definitely hopeless. However, people
look down on the peculiar quantum nonlocal influence, in fact, it
not only is independent of the limitation of present quantum
theory on superluminal communication, but also rejects special
relativity to some extent\cite{Hardy}, here we will demonstrate
that the description about quantum nonlocal influence needs a
preferred Lorentz frame, and the quantum superluminal
communication based on such quantum nonlocal influence, if exists,
does not result in the causal loop, this undoubtedly opens the
first door to superluminal communication.

At first, Hardy's theorem\cite{Hardy} first states that any
dynamical theory describing the quantum nonlocal process, in which
the predictions of the theory agree with those of ordinary quantum
theory, must have a preferred Lorentz frame, and the description
about the quantum nonlocal influence is no longer independent of
the selection of inertial frame, this evidently breaks the first
assumption of special relativity, which asserts that the
description of any physical process is independent of the
selection of inertial frame. But in Hardy's proof he presupposed
that the collapse process happens simultaneously in all observing
inertial frames or there is no backward causality in quantum
systems, which validity is still not clear, this weakens the
strength of his conclusion.

Then, Percival extended Hardy's theorem, he gave a different
derivation based on classical links between two Bell experiments
in different experimental inertial
frames\cite{Percival1,Percival2,Percival3}, which is called the
double Bell experiment, and his proof is independent of any
assumptions about causality in the quantum domain, thus it is just
the quantum nonlocal influence itself that requires the dynamical
theory about it must have a preferred Lorentz frame, or there will
exist the forbidden causal loops in the systems with classical
inputs and outputs.

On the other hand, Suarez's analysis about
multisimultaneity\cite{Suarez1,Suarez2} has also indicated that
the description about the causal orders of the nonlocal
correlating events essentially needs a preferred Lorentz frame,
although he didn't realize this fact himself, in fact, his elegant
one Bell experiment involving 2-after impacts will also generate
the forbidden logical causal loop if we assume that no preferred
Lorentz frame exists, or the quantum nonlocal influence happens
simultaneously in all experimental frames, since as to these two
space-like classical events in the experiment, each is the cause
of the other, and this is evidently a logical contradiction. Thus
Suarez's one Bell experiment also demonstrates that there must
exist a preferred Lorentz frame in order to consistently describe
the quantum nonlocal process.

Now, all the above demonstrations have clearly indicated that the
consistent description about the quantum nonlocal influence needs
a preferred Lorentz frame, in which all quantum nonlocal
influences are simultaneous, and the causal relation between the
correlating quantum nonlocal events are exclusively determined,
then all quantum nonlocal influences will be no longer
simultaneous in other inertial frames according to Lorentz
transformations, in fact, in these frames, the quantum nonlocal
influence, or quantum simultaneous communication if exists, will
proceed forward in time along one direction in space, and proceed
backward in time along the contrary direction in space, the causal
relations between these correlating quantum nonlocal events in
these frames will no longer relate directly to their time orders,
and be only determined by their time orders in the preferred
Lorentz frame, then it is evident that there will no longer exist
any causal loops for the quantum nonlocal influence and possible
quantum simultaneous communication based on such quantum nonlocal
influence, since the causal relations of the correlating quantum
nonlocal events are exclusively determined by their time orders in
the preferred Lorentz frame, and causes always come before
effects.

At last, I will give a simpler apagogical demonstration, namely if
the quantum simultaneous communication based on the quantum
nonlocal influence leads to the forbidden causal loops, then the
quantum nonlocal influence itself will also lead to the forbidden
causal loops, the reason is simple, since in Percival's double
Bell experiment\cite{Percival1,Percival2,Percival3}, if we devise
the experimental settings in order that the quantum simultaneous
communication based on the quantum nonlocal influence leads to the
forbidden causal loops with certainty, then the quantum nonlocal
influence itself will also lead to the forbidden causal loops with
a nonzero probability, as Percival has minutely
demonstrated\cite{Percival1,Percival2,Percival3}, this is not
permitted either, thus we again get the conclusion, namely quantum
superluminal communication, if exists, does not result in the
causal loop.

\vskip 1cm \noindent Acknowledgments \vskip .5cm Thanks for
helpful discussions with A.Suarez ( Center for Quantum Philosophy
), Dr S.X.Yu ( Institute Of Theoretical Physics, Academia Sinica
).

\end{document}